\documentclass[onecolumn,preprintnumbers,amsmath,aps]{revtex4-1}
\usepackage{graphicx}
\usepackage{bm}
\usepackage{subfigure}
\usepackage{color}
\usepackage{amssymb}
\usepackage[rgb]{xcolor}
\usepackage{soul}

\begin{document}


\title{Entropic signatures of market response under concentrated policy communication}


\author{Ewa A. Drzazga-Szcz{\c{e}}{\'s}niak$^{1}$}
\author{Rishabh Gupta$^{2,3}$}
\author{Adam Z. Kaczmarek$^{4}$}
\author{Jakub T. Gnyp$^{5,6}$}
\author{Marcin W. Jarosik$^{1}$}
\author{R{\'o}{\.z}a Walig{\'o}ra$^{5}$}
\author{Marta Kielak$^{4}$}
\author{Shivam Gupta$^{4}$}
\author{Agata Gurzyńska$^{6}$}
\author{Johann Gil$^{4,7}$}
\author{Piotr Szczepanik$^{5}$}
\author{J{\'o}zefa Kielak$^{5}$}
\author{Dominik Szcz{\c{e}}{\'s}niak$^{4}$}
\email{d.szczesniak@ujd.edu.pl}


\affiliation{
${^1}$Department of Physics, Faculty of Production Engineering and Materials Technology, Cz{\c{e}}stochowa University of Technology, 19 Armii Krajowej Ave., 42200 Cz{\c{e}}stochowa, Poland\\
${^2}$Department of Chemistry, Purdue University, 560 Oval Dr., 47907 West Lafayette, Indiana, United States\\
${^3}$Department of Electrical and Computer Engineering, North Carolina State University, Raliegh, North Carolina 27606, United States\\
${^4}$Institute of Physics, Faculty of Science and Technology, Jan D{\l}ugosz University in Cz{\c{e}}stochowa, 13/15 Armii Krajowej Ave., 42200 Cz{\c{e}}stochowa, Poland\\
${^5}$Institute of Pricing and Market Analysis, Analitico, 186 Francuska Str., 40507 Katowice, Poland\\
${^6}$Condensed Matter Spectroscopy Division, Faculty of Mathematics, Physics and Informatics, University of Gda{\'n}sk, Wita Stwosza 57 Str., 80308 Gda{\'n}sk, Poland\\
${^7}$Institute for Molecules and Materials UMR 6283, Le Mans University, Ave. Olivier Messiaen, 72085 Le Mans, France}


\date{\today}


\begin{abstract}
The first 100 days of Donald Trump second presidential term (January~20th - April~30th,~2025) featured policy actions with potential market repercussions, constituting a well-suited case study of a concentrated policy scenario. Here, we provide a first look at this period, rooted in the information theory, by analyzing major stock indices across the Americas, Europe as well as Asia and Oceania. Our approach jointly examines dispersion (standard deviation) and information complexity (entropy), but also employs a sliding window cumulative entropy to localize extreme events. We find a notable decoupling between the first two measures, indicating that entropy is not merely a proxy for amplitude but reflects the diversity of populated outcomes. As such, they allow us to capture both market volatility and narrative constraints, signaling large and coherent moves driven by policy changes. In turn, the cumulative entropy is found to notably increase during regional episodes with high information density, providing effective signatures of such events. We argue that the obtained results indicate short-term globally coupled, yet regionally modulated, market impacts with clear connection to introduced policies. In what follows, the presented entropic framework emerges as an efficient complement to standard methods for characterizing markets under turbulent conditions, with potential to enhance forecasting strategies such as the stochastic modeling.
\end{abstract}


\maketitle

\section{Introduction}
\label{sec:intro}

The first 100 days of Donald Trump second presidential term became one of the most widely discussed early presidencies in recent history \cite{stimpson2025}. Even before taking office, Donald Trump, as a candidate, promised reforms that may be viewed by some as courageous or even radical \cite{dombrowski2024,bremmer2024}. Indeed, this initial period of Trump presidency was marked by a wave of key decisions, including announcements of large-scale industrial projects, interventions related to ongoing military conflicts, but most importantly tariffs \cite{arato2025,satoru2025,rickard2025}. At first sight, all these actions appear to align with the famous phrase {\it make America great again}, suggesting a potential drive to strengthen the leadership of the United States (U.S.) on the world stage. However, it remains an open question whether empirical evidence supports this narrative, or whether these policies deliver tangible benefits to the U.S. economy and the global system more broadly. In this context, the condition of global financial and trade markets may provide a particularly informative benchmark \cite{eldomiaty2024}. We also note, the discussed period can serve as an effective testbed for financial analysis methods within a highly disrupted market environment characterized by concentrated policy communication, thereby enabling an assessment of their robustness and reliability.

Here, we attempt to address these aspects, by examining a broad cross-section of major equity indices across the Americas, Europe as well as Asia and Oceania. This is done over the 100-day window that follows President Trump inauguration on January~20th,~2025. When necessary, we contrast these findings against the analogous 100-day period preceding the inauguration. The uniqueness of our approach lies in the direct use of information theory measures to quantify market disturbance, instead of just discussing dispersion as summarized by the standard deviation of returns. In particular, the information complexity is characterized here by the Shannon entropy \cite{shannon1948,lahmiri2017disturbances,delgado2019,delgado2021,shternshis2022,rodriguez2022,das2024complexity}. While dispersion quantifies the magnitude of price movements \cite{stigler1970dispersion}, entropy measures the degree of unpredictability and disorder within the return series \cite{pele2017information, drzazga2023}. Hence, a market with high entropy is characterized by a more uniform distribution of outcomes, indicating randomness and informational efficiency \cite{ahn2019stock}. In contrast, low entropy suggests a more ordered, predictable, or constrained state, where certain return outcomes are more favored than the others \cite{papla2024entropy}. Thus, entropy can be regarded as a viable tool for measuring market efficiency \cite{shternshis2022}, volatility or even serving as a possible marker for extreme events \cite{sheraz2015,drzazga2023,das2024complexity}. Together, these statistics allow us to distinguish between markets that are merely volatile and markets that are volatile yet narratively constrained.

Motivated by the above, as well as by the prominence of policy communications and trade actions in early 2025, we posit that informational complexity can provide new insights into a turbulent market behavior. In particular, two intersecting rationales are advanced to develop this argument. Our first hypothesis is that when a small number of salient narratives repeatedly channel market reactions into similar configurations, the distribution of returns becomes more concentrated over a few recurrent outcomes even if amplitudes remain large. In such conditions, entropy may compress while volatility is high, a regime we refer to as structured volatility. This motivates measuring both standard deviation and Shannon entropy on the same windows and reading them jointly rather than treating entropy as a proxy for amplitude. Taken together, these considerations represent a development beyond the idea of entropy as merely an alternative measure of volatility previously presented in \cite{sheraz2015,drzazga2023}. A second hypothesis directly follows this and concerns the timing of market dislocations. If extreme episodes are brief and dense in information, then evaluating entropy on short windows anchored to a moving reference and aggregating it across expanding horizons should produce sharp, ramp-like profiles around the onset of those episodes. We therefore adopt a sliding, expanding-window cumulative entropy construction designed to localize (i) when informational complexity builds most rapidly, (ii) how intensely it accumulates, and (iii) how persistent the elevated state is, without prespecifying event windows or imposing a parametric shock model \cite{drzazga2025}. That is, the cumulative measure is benchmarked here against not only turbulent but potentially extremely noisy data.

Interpreting these patterns across markets requires acknowledging that transmission channels are not uniform. In settings with tighter policy, trade, or financial linkages, one would expect entropy dynamics to move more nearly in step and cumulative entropy profiles to cluster in time, consistent with a more event-driven and globally coupled regime. By contrast, where local factors like market microstructure play a larger role, the same tools should reveal a weaker synchronicity and more variations in the pace or shape of the cumulative entropy buildup. The aim, therefore, is not to argue for uniformly calm or uniformly turbulent markets, but to show how volatility-based dispersion metrics and information-theoretic measures can be read together. The former serves to flag possible decoupling between risk and information load, the latter to help locate and describe the episodes that give rise to it. In this context, the paper is organized as follows: Section~\ref{sec:dataprep} details data preprocessing, Section~\ref{sec:methods} formalizes the methodology, Section~\ref{sec:volatility} presents side-by-side entropy and standard deviation comparisons across indices and regions, Section~\ref{sec:extreme} analyzes extreme events using selected cumulative entropy panels, and Section~\ref{sec:conclusion} summarizes findings as well as outlines implications and avenues for longer-horizon assessment.

\section{Data preprocessing}
\label{sec:dataprep}

\begin{table*}[ht]
\renewcommand{\arraystretch}{1.9} 
\centering
\setlength{\tabcolsep}{3.8pt} 
\caption{Market stock indices analyzed in the current study along with their relevance to the U.S. policy developments. The stock indices are provided together with the corresponding regions, names, symbols/abbreviations, markets as well as frequency and data providers (in square brackets).}
\begin{tabular}{|c|p{2.0cm}|p{1.5cm}|p{1.5cm}|p{6.4cm}|p{2.7cm}|}
\hline
\textbf{Region} & \textbf{Name index} & \textbf{Symbol} & \textbf{Market} & \textbf{Relevance} & \textbf{Source}\\
\hline
\hline
Americas 
        & S\&P 500 & SPX & U.S. & Broad representation of U.S. market; directly impacted by domestic policies. & 5 min [STQ] \newline daily [INV]  \\
         & DJIA & DJI & U.S. & Covers major industrial sectors; sensitive to economic policy.  & 5 min [STQ] \newline daily [EOD] \\
         & NASDAQ \newline Composite & NDQ & U.S. & Heavily weighted in technology and growth companies. & 5 min [STQ] \newline daily [INV] \\
         & Bovespa & BVP & Brazil & Largest Latin American market; affected by trade and commodity trends. & 5 min [STQ] \newline daily [INV]\\
         & S\&P/TSX \newline Composite & TSX & Canada & Strong trade ties with the U.S.; energy and materials focus. & 5 min [STQ] \newline daily [INV]\\
\hline
Europe   
        & Euro Stoxx 50 & EUS.IDX & Eurozone & Represents major blue-chip companies in the EU; reflects overall sentiment. & 5 min [DUK] \newline daily [DUK]\\
         & FTSE 100 & GBR.IDX & UK & Post-Brexit UK maintains close economic relations with the U.S. & 5 min [BLU] \newline daily [DUK]\\
         & DAX & DAX & Germany & Germany’s industrial sector sensitive to global trade policies. & 5 min [STQ] \newline daily [INV]\\
         & WIG20 & WIG20 & Poland & Central/Eastern European representation and emerging market. & 5 min [STQ, DUK] \newline daily [STQ, DUK]\\
\hline
Asia     
         & Nikkei 225 & NKX & Japan & Export-driven economy; closely aligned with U.S. policy. & 5 min [STQ] \newline daily [INV]\\
         & Shanghai \newline Composite & SHC & China & Key index for monitoring U.S.-China trade and economic tensions. & 5 min [STQ] \newline daily [INV]\\
         & Hang Seng \newline Index & HSI & Hong Kong & Mix of local and Chinese firms; reacts to U.S.-China geopolitical dynamics. & 5 min [STQ] \newline daily [INV]\\
         & Nifty 50 & NSEI & India & Reflects India’s growing economy and ties with the U.S. in tech and defense. & 5 min [EOD] \newline daily [EOD]\\
\hline
Oceania  
        & S\&P/ASX 200 & AUS.IDX & Australia & Strong exposure to both Chinese and U.S. economic policies.  & 5 min[DUK] \newline daily [DUK]\\
        & NZX 50 & NZ50& New Zealand & Smaller but regionally important; trade-sensitive. & 5 min [STQ] \newline daily [INV]\\
\hline \hline
\end{tabular}
\label{tab01}
\end{table*}

Taking into consideration the broad range of influence of the American economy in the current international order, a number of market indices were chosen for analysis from each major geographical region, as specified in Table \ref{tab01}. Political and economic relations between various markets were carefully examined to highlight key linkages between countries and specific industries. Thus, instead of simply choosing indices from countries with the highest GDP, indices such as the Nasdaq Composite, Euro Stoxx 50, and Shanghai Composite were included in the analysis.

The basic statistics and the market volatility were analyzed based on the daily data, which is sufficiently dense for such purpose. However, more information-rich data was required for the cumulative entropy calculations. In this case, using 5-minute data was a pragmatic choice, leveraging data accessibility and its granularity. The market data was collected from several sources to obtain full coverage of the considered time periods. The employed data providers, abbreviated as STQ (stooq.com), INV (investing.com), EOD (eodhd.com), DUK (dukascopy.com), BLU (bluecapitaltrading.com), are listed in Table \ref{tab01} with respect to each considered stock index. As already mentioned, for comparison of the empirical statistic distributions of the data after Trump's inauguration on January~20th,~2025 the data from a hundred days from before the inauguration was gathered as well. However, since the period from before the inauguration is treated as a series of referential entropy values, the daily data was again sufficient. Even though several extreme events may be found in this period, their frequency was significantly smaller than in the data after the inauguration, as it will be explicitly shown in the next sections.

For entropy calculations a unified target data format was created and all datasets were converted to it. This standardization process mainly consisted of converting various timestamp and date-time formats, depending on the data source, into one. Formats YYYY-MM-DD for the daily data and YYYY-MM-DD HH-MM-SS for the 5-minute data were selected. Few missing values were identified during market open hours, of no statistical significance. For several datasets, the data points for the closed market hours, usually multiple copies of the last open market data point, were identified and deleted.

Afterwards, for the daily data from before the inauguration and the intraday data from after the inauguration, aggregated into daily data, nominal intraday returns and log returns were computed. Additionally, mean and variance values, quartiles, Pearson kurtosis and adjusted Fisher-Pearson standardized moment coefficients for skewness were calculated, as shown in the Tables \ref{tab02} and \ref{tab03} in the Appendix. 

For convenience, the data were collected in an in-house-developed relational database implemented using the MariaDB (a fork of MySQL) SQL dialect. This setup enabled uniform access to standardized data for the Python-based data analysis modules also created internally for this study. As a result, the complete database and preprocessing pipeline can be readily extended and reused in future work.

\section{Methodology}
\label{sec:methods}

First, we attempt to formalize the two main information measures used throughout the paper. We begin with the Shannon entropy of binned return distributions, which is employed here as an estimate of informational complexity at a given time horizon. Second, we consider a cumulative entropy with a sliding window, which localizes the timing and intensity of short but information-dense events.

\subsection*{Shannon entropy}

The information encapsulated within a financial data (informational complexity) is quantified here using the Shannon entropy, as computed on interval (binned) return distributions. Note that entropy, in its original formulation, was applied to a variety of subfields and problems within thermodynamics \cite{kumar2025, kaczmarek2024}. Shannon entropy draws inspiration from this concept but has a fundamentally different interpretation. It does not describe physical randomness, but rather the average number of binary questions required to decode a message \cite{drzazga2023}. In detail, we assume that a return series is considered for a given index and analysis window (e.g., a day or a block of consecutive days). For such series, we choose a binning method and partition the returns into $m$ contiguous intervals or bins. In our analysis, we use the so-called Velleman formula for the partitioning purpose \cite{dougan2010}. Next, $p_i$ denotes empirical probability that a return observed in a given window falls into bin $i$ ($i=1,\dots,m$), so that $\sum_{i=1}^m p_i=1$. The corresponding Shannon entropy for a given window is then \cite{drzazga2025}:
\begin{equation}
\label{eq01}
H=-\sum_{i=1}^{m} p_i {\rm log} p_i.
\end{equation}
In this form, Eq. (\ref{eq01}) measures how diffused or concentrated the distribution of returns is across the intervals or bins of choice. In contrast, the standard deviation, a conventional volatility measure, summarizes the amplitude of fluctuations, regardless of whether those fluctuations are structured or individual. To ensure comparability across windows and indices, the number of bins used in the computations is fixed for all calculations conducted here, according to the mentioned Velleman formula.

In what follows, the calculated entropy captures the diversity of realized return patterns present in a given window. In other words, the value of entropy is high when many bins carry a comparable probability mass, indicating numerous distinct patterns, and low when the mass concentrates in fewer bins, reflecting a dominance of repeating patterns. Because it is agnostic to the metric scale of returns, entropy is complementary to measures based on dispersion, like the already mentioned standard deviation. Their joint analysis distinguishes between merely volatile markets and those that are volatile yet narratively constrained, in which large moves are repeatedly channeled into a narrower repertoire of outcomes.

\subsection*{Cumulative entropy}

It was previously reported that entropy can be also used to detect extreme events in financial data \cite{drzazga2023, drzazga2025}. In particular, based on the introduced above Shannon entropy, the so-called cumulative entropy can be formally defined \cite{drzazga2025}. For this purpose, let us consider a discrete time series that takes values from a stochastic process, with times $T \subset \mathbb{N}_0$, where $(X_t)_{t \in T}$ is a family of real random variables of returns of a given financial instrument, e.g. an index or a stock. While each random variable in the mentioned family $(X_t)_{t \in T}$ follows a true underlying distribution $p_{\text{true}}$, the time series $x_t$ represents only a finite, time-restricted collection of realizations, described instead by an empirical probability distribution $p_{\text{emp}}$. The empirical distribution converges almost surely to the underlying probability distribution in the limit where the cardinality of the subset tends to infinity, as proved by Glivenko and Cantelli \cite{glivenko1933, cantelli1933}.

With extreme events occurring in the discussed time series, its empirical distribution will protrude from the true, underlying one for the non-disturbed family $(X_t)_{t \in T}$. However, gathering enough data to fit and compare distributions with a satisfying accuracy is not computationally effective. Thus, another method is called for, with a smaller computational cost and a greater transparency. One such measure is the mentioned cumulative entropy, representing the amount of informations needed to describe the system with the extreme events occurring in it.

To formally define this measure, we first note that financial time series of returns are characterized by numerous distinct values with a little to no repetitions and leaving the values unrounded is appropriate. Therefore, it is impractical to compute any entropy for probabilities of specific empirical values of returns. Instead, one could group the returns into $n$ evenly spaced intervals of width $\Delta x$, where $n$ is given again by the Velleman formula \cite{dougan2010}. Probabilities of values $x_t$ of the time series belonging to one or an other interval form in turn the interval probability distribution $p_{int}$. Thus, instead of point probabilities  $p_{emp}$ associated with specific return values $x'$, we consider {\it interval} probabilities $p_{int}$ corresponding to returns defined over return intervals $[x_i,x_{i+\Delta x}]$. However, this does not yet contain distinct informations on any extreme values in the time series as a whole. Therefore, instead of grouping all values of $(x_t)_{t \in T}$ for the full time span $T$, a series of entropy values could be calculated for multiple decreasing sequences of time windows:
\begin{equation}
\label{eq02}
    T_0^{(1)} \subset T_1^{(1)} \subset \dots T_m^{(1)},\dots,T_0^{(M)} \subset T_1^{(M)} \subset \dots T_m^{(M)},
\end{equation}
where $T_m^{(1)},\dots,T_m^{(M)}$ is a sequence of partially overlapping subsets of the full time span $T$. Each such time window from each of the increasing sequences contains more discrete times by an increment of a fixed number $\Delta t$. 

Thus, for each sequence $T_0 \subset T_1 \subset \dots T_m$ multiple entropy values are considered:
\begin{eqnarray}
    \nonumber
    H_0 &= -\sum_{i=1}^n p_i^{(0)} \ln p_i^{(0)}, \\ \nonumber
    H_1 &= -\sum_{i=1}^n p_i^{(1)} \ln p_i^{(1)}, \\ \nonumber
    &\vdots \\
    H_m &= -\sum_{i=1}^n p_i^{(m)} \ln p_i^{(m)}, \nonumber
\end{eqnarray}
where $p_i^{(k)}$ denotes the fraction of return observations within the time window $T_k$ that fall into the interval $[x_i, x_i+\Delta x]$. In what follows, the entropy $H_m$ may be denoted in short as:
\begin{equation}
\label{eq03}
    H_m = -\sum_{i=1}^n p_i^{(m)} \ln p_i^{(m)},
\end{equation}
meaning that the probability of some return value $x'$ belonging to an $i$-th interval of the time window $T_m$ is considered. Equation~(\ref{eq03}) defines a cumulative entropy in the sense that it encapsulates how the entropy of a time series changes as time windows of increasing cardinality are considered. Such windows may contain an extreme event, which can drastically alter the interval probability distribution for some time windows in comparison to the rest. Therefore, it is best to consider a cumulative entropy spectrum where each cumulative entropy spans over different time windows from a sequence $T_0 \subset T_1 \subset \dots T_m$.

\section{Market Volatility}
\label{sec:volatility}

We begin our analysis by studying the market behavior across two distinct periods of 100 days, that precede and follow the presidential inauguration on January~20th,~2025, respectively. In particular, this is done in a twofold manner, namely by employing the standard deviation of returns to quantify dispersion and via Shannon entropy to measure the informational complexity. Note that these calculations are conducted for the daily data sets only, which nevertheless provide sufficient sample density to populate bins defined previously within Velleman formula.

\begin{figure}[ht!]
\includegraphics[width=0.5\columnwidth]{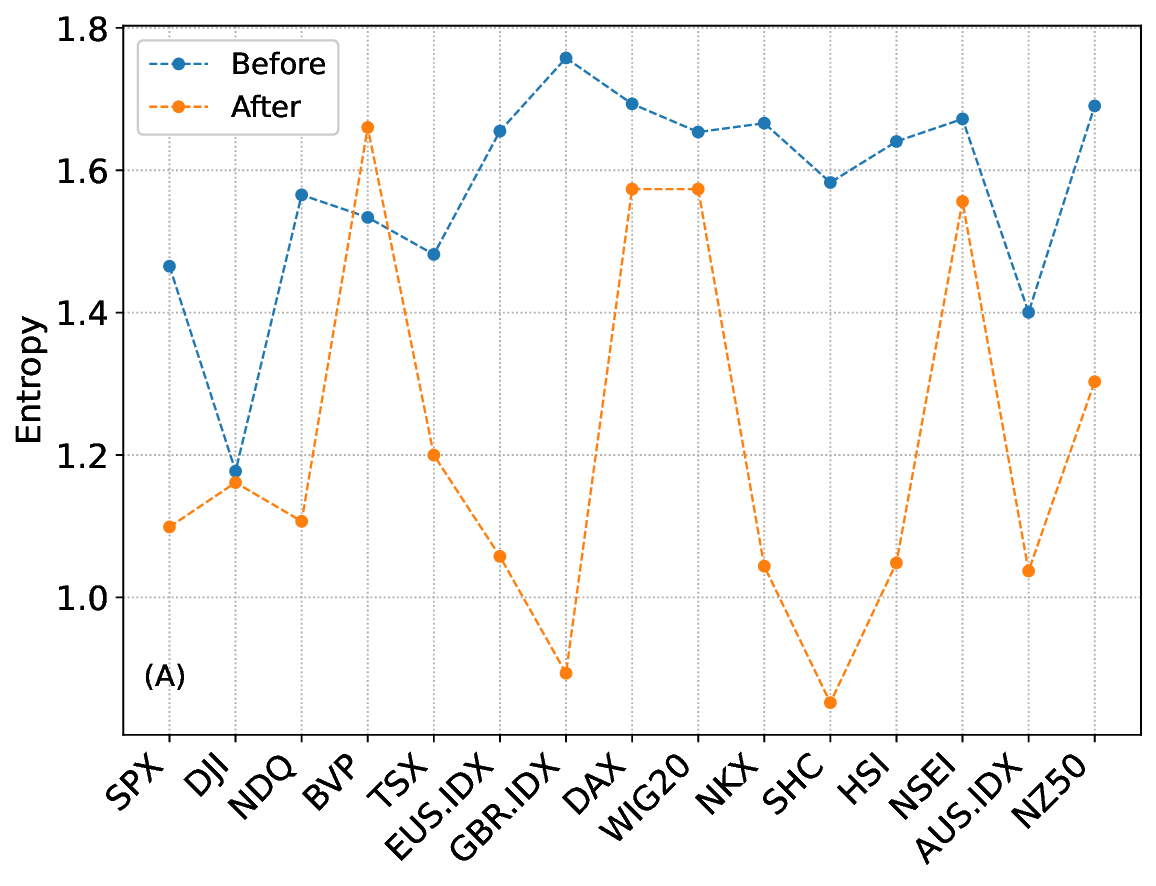}\includegraphics[width=0.5\columnwidth]{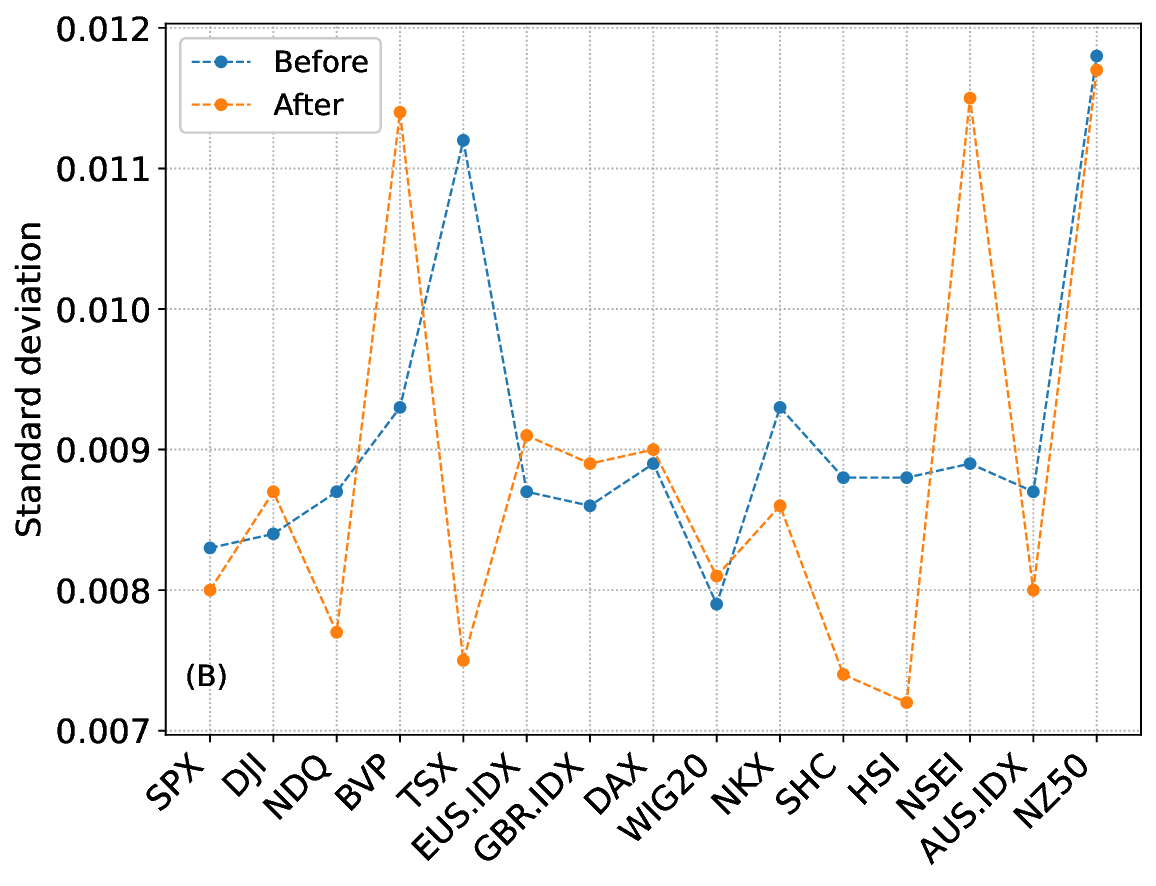}
\caption{Entropy and standard deviation for each considered stock index over symmetric 100-day windows bracketing presidential inauguration on January~20th,~2025. The pre- and post-inauguration periods are labeled as {\it before} and {\it after} results, respectively.}
\label{fig01}
\end{figure}

The corresponding results, for both metrics across all considered indices (see Table \ref{tab01}), are depicted in Figure \ref{fig01}. Interestingly, at first glance, these results appear counterintuitive, showing that entropy frequently declines in the post-inauguration period whereas standard deviation remains comparable or higher than its counterparts from before presidential inauguration. In other words, the results do not present heightened entropy as it may be intuitively expected in the case of large and/or frequent changes of the return value. However, after deeper inspection, it is argued here that this unexpected observation can be explained by stating that entropy is not simply a proxy for amplitude but in fact it is a measure of how many distinct patterns of outcomes become populated. For example, if market reactions are channeled by a small set of dominant narratives (like analyzed recurring policy themes), price changes can be large but coherent. This is to say, the probability shifts toward a more limited set of return intervals that emerge around similar announcements. In such a regime, entropy falls even when standard deviation stays elevated, as observed in Figure \ref{fig01}. In what follows, dispersion and randomness appear somewhat distinct, meaning volatility quantifies how far prices move whereas entropy measures how spread out the behavior of those moves is. This is to say, on days influenced by the news related to some policy, returns should tend to cluster into a small number of bins, reducing entropy. In contrast, on days with more diffuse or cross-cutting flows, the probability mass should disperse across more bins, raising the entropy measure. Figure \ref{fig01} makes clear that the post-inauguration period often belongs in the former regime. Importantly, this argumentation seems to also explain conceptual and empirical differences between standard deviation and entropy, observed previously in \cite{sheraz2015,drzazga2023}.

In addition to the above, Figure \ref{fig01} presents contrasts between considered regions. In particular, several indices of developed markets (in both Europe and Asia) exhibit well-visible declines in entropy along with a flat or a moderately higher dispersion. In our opinion, this is consistent with the idea that trade and policy channels generate coherent signals that markets repeatedly price in. Remaining markets show a weaker entropy compression or even the opposite sign values, indicating that local characteristics, commodity sensitivities, or domestic factors can dilute this coherence effect. The volatility there may reflect a broader mixture of unrelated shocks or extreme events, preserving a relatively high entropy. These results thus support a simple observation, namely where the post-inauguration news flow is directed through a few dominant channels, entropy falls. In contrast, where multiple unrelated forces compete, entropy is more resilient.

\begin{figure}[ht!]
\includegraphics[width=0.5\columnwidth]{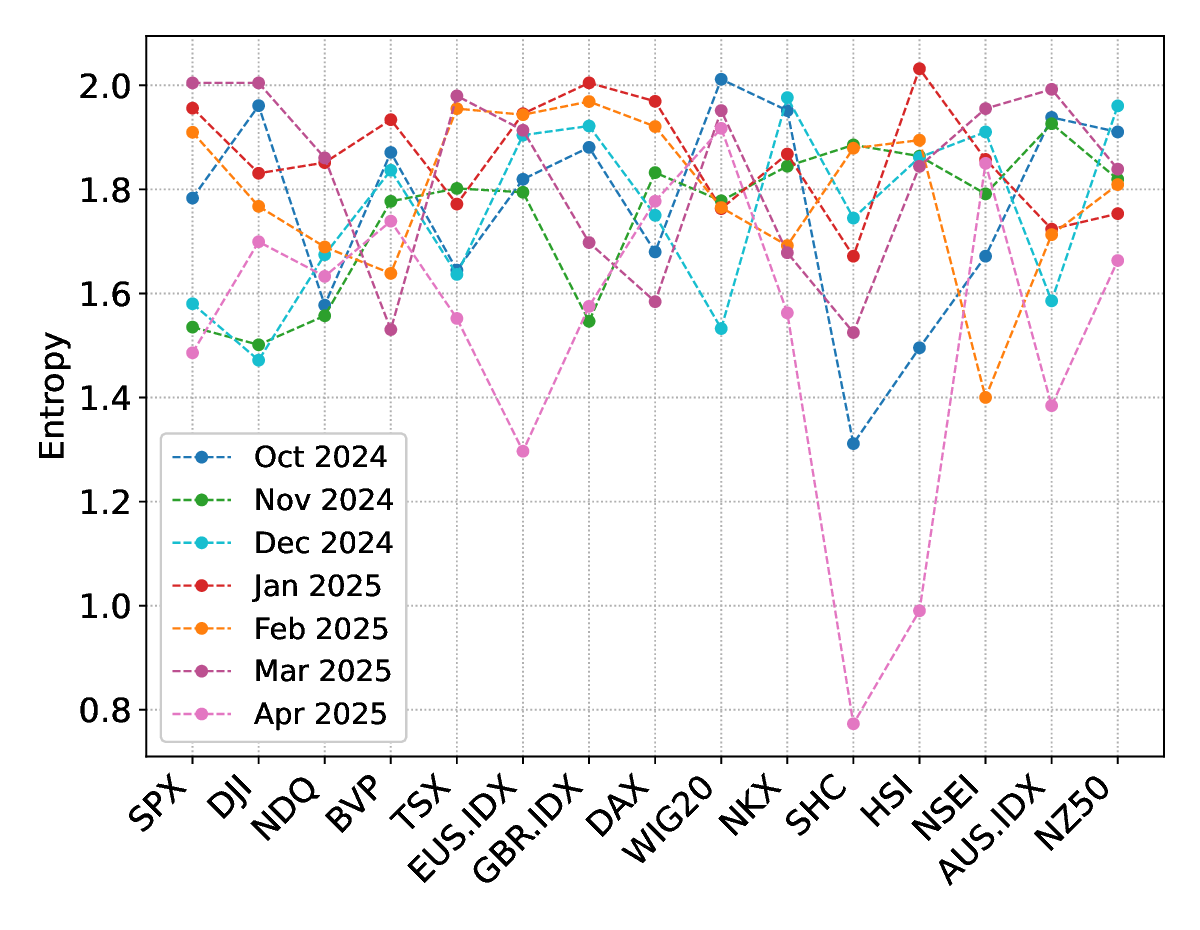}
\caption{Monthly Shannon entropy profile from October~2024 through April~2025 for each considered stock index. The results are bridging the 100-day pre-inauguration baseline and the post-inauguration window.}
\label{fig02}
\end{figure}

The already discussed findings are supplemented by the results on monthly entropy, as given in Figure \ref{fig02}, showing that changes in entropy are occasional rather than uniform. Months marked by concentrated and consistent news exhibit deeper channels in entropy, whereas months with more mixed developments show a partial broadening of the analyzed measure. Such behavior explains why contrasts between results before and after presidential inauguration (see Figure \ref{fig01}) can remain strong on average, even during weeks when entropy temporarily rebounds. Obviously, the coherence in the news narrative is inherently not constant, hence the observed sensitiveness of entropy measures to the shifts in informational structure may be particularly valuable in the corresponding analysis.

\begin{figure}[ht!]
\includegraphics[width=0.5\columnwidth]{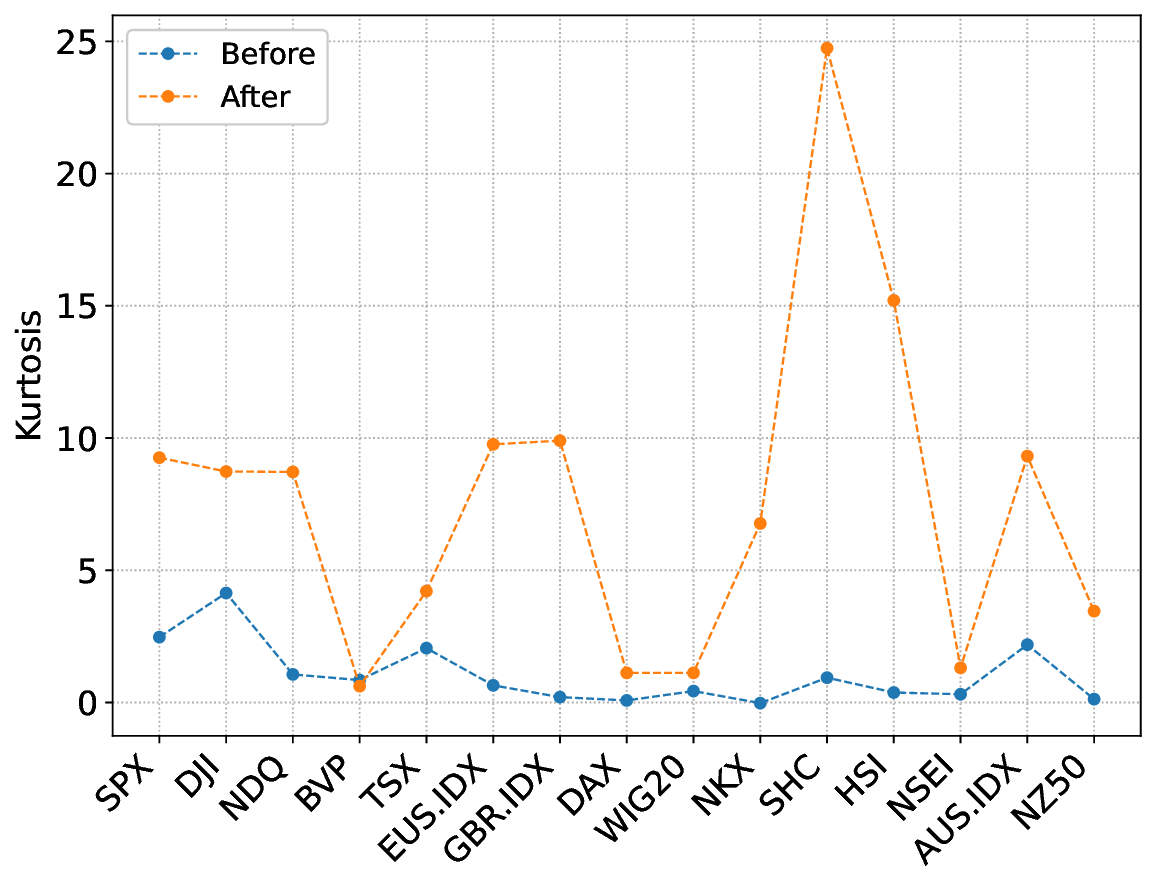}
\caption{Kurtosis of 5 minutes returns for each considered stock index. The pre- and post-inauguration periods are labeled as {\it before} and {\it after} results, respectively.}
\label{fig03}
\end{figure}

The alignment between this narrative and distributional findings is reinforced by the calculated kurtosis. In particular, Figure \ref{fig03} shows an elevated kurtosis at higher sampling frequencies, suggesting a heavy-tailed return behavior that is compatible with a lower entropy. Such heavy tails indicate that when markets move, they can move far. On the other hand, the lower entropy describes why and how their moves are organized around a smaller set of repeatable patterns. Together, Figure \ref{fig01} and Figure \ref{fig02} show that the early 2025 regime is best characterized not as undisturbed, but as volatile yet narratively constrained with large price jumps occurring within a relatively ordered information landscape.

Interestingly, the joint behavior displayed in Figures \ref{fig01} and \ref{fig02} may also lead to a more fundamental interpretation of entropy that goes well beyond presented descriptive market diagnostics. In this respect, the key empirical observation is that across many considered indices the post-inauguration period exhibits a preserved or an increased dispersion along with the reduced entropy. This observation implies that, despite large price movements, the set of corresponding returns becomes somewhat more restricted and repetitive. We argue that such a concentration of probability masses signifies an increase in deterministic content. In detail, outcomes are no longer sampled freely from a broad state space but are repeatedly drawn from a smaller, structured subset. Figure \ref{fig02} further confirms that this entropy behavior is not a temporary noise but persists over some monthly regimes, indicating stable constraints of market dynamics rather than incidental fluctuations.

Based on the above, it can be argued that the observed reduction in accessible information states reinforces treating entropy as a quantitative measure of the deterministic structure embedded within an otherwise stochastic process. This observation is particularly important from a forecasting perspective, since the results shown in Figures \ref{fig01} and \ref{fig02} establish entropy as a natural control variable. This is to say, when entropy is low, the system departs from a random and diffusion dominated regime, often assumed by some stochastic processes used in financial analysis. Under such conditions, a stochastic forecast that ignores entropy is systematically overstating randomness. Importantly, this observation is consistent with earlier results reported by some of us on entropy-based corrections to the standard geometric Brownian motion framework \cite{gupta2024}. Accordingly, the regimes identified in Figures \ref{fig01} and \ref{fig02} describe additional conditions under which reduced entropy signals that future trajectories should be constrained toward distributions reflecting a heightened deterministic content rather than an unconstrained log-normal diffusion.

\section{Extreme Events}
\label{sec:extreme}

Next, we interpret the cumulative entropy as a model-agnostic lens that converts local changes in a return distribution into a time specific signal or a signature. In practice, we compute entropy on a fixed bin return distribution within a short window and then slide this window forward in time to obtain a so-called cumulative entropy. When a market approaches an extreme event or a shock, the local distribution of returns is reconfigured and the probability mass shifts among a small number of bins so the cumulative form of entropy increases visibly. As a result, the mentioned increase in entropy often appears exponential-like, producing a clean and visually notable signature.

\begin{figure}[ht!]
\includegraphics[width=0.47\columnwidth]{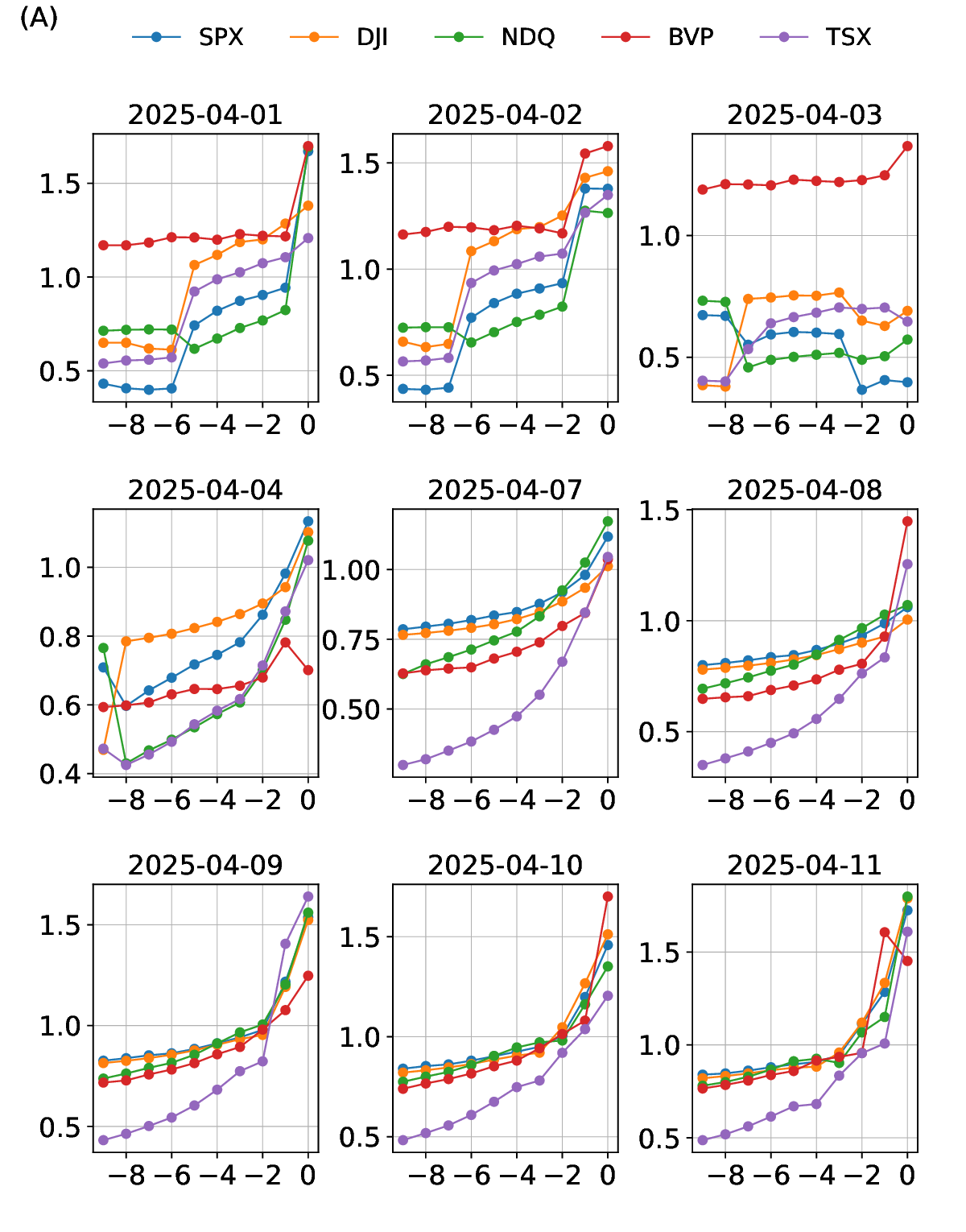}\includegraphics[width=0.47\columnwidth]{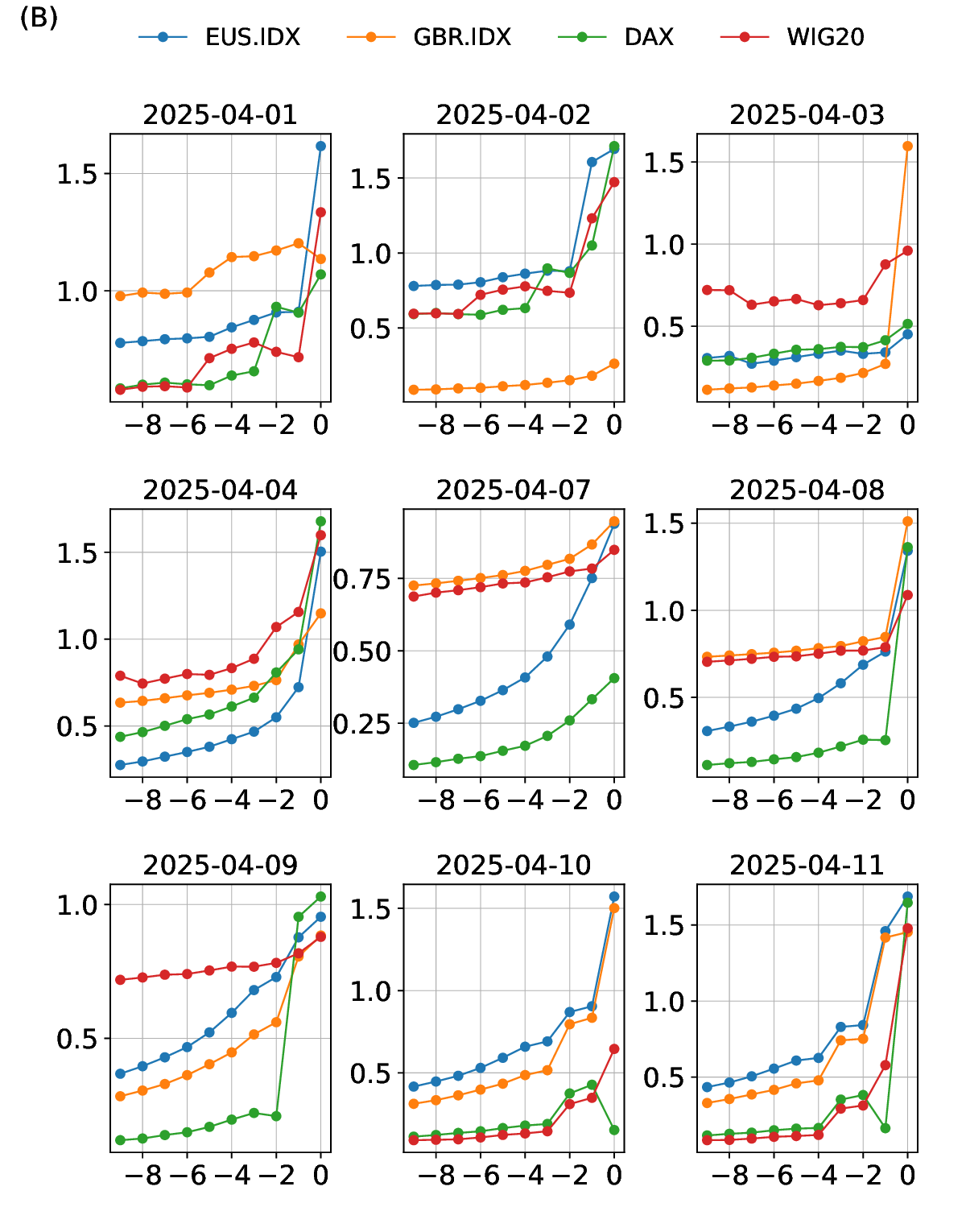}\\ \includegraphics[width=0.47\columnwidth]{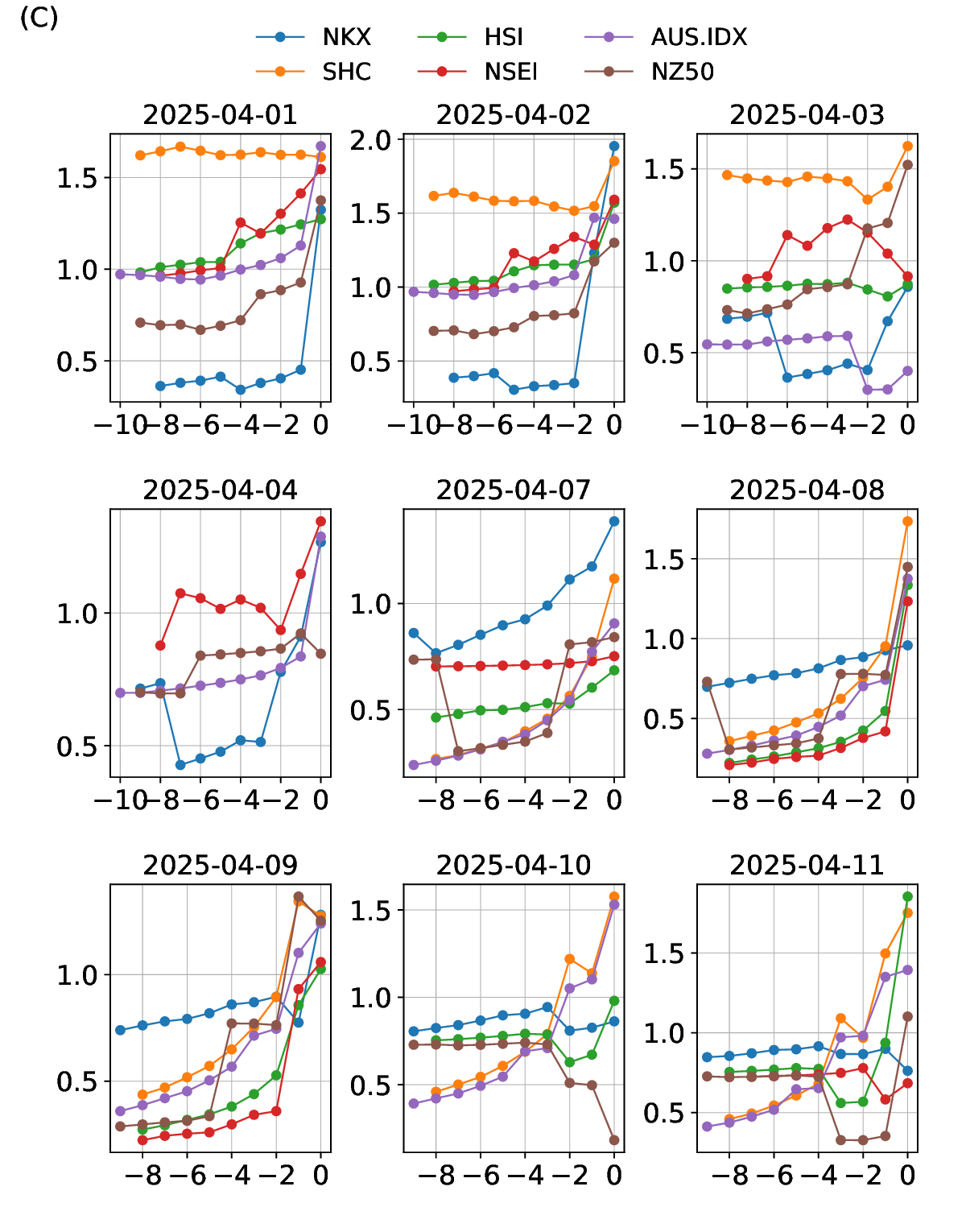}
\caption{Cumulative entropy for April~1–11, 2025 across representative Americas (A), European (B) as well as Asian and Oceania (C) stock indices.}
\label{fig04}
\end{figure}

Amongst the several signatures of extreme events, as presented in the Appendix B, the results for April~1–11 stand out for the Americas, Europe, as well as Asia and Oceania panels (see Figures \ref{fig04} (A)-(C)), illustrating the occurrence of tarrifs introduction. In particular, each subfigure shows a pronounced increase in cumulative entropy that begins near the start of April and reaches a maximum value within a few sessions, before either stabilizing at a higher level or gradually unwinding. In our opinions, the common structure across regions is informative, meaning the onset is sharp (indicating a mentioned reconfiguration of probability mass), the slope during the rise is steep (indicating a dense clustering of similarly patterned market responses), and the decay or plateau afterwards varies by market (indicating differences in local responses and narratives). This pattern supports an unified interpretation in which a small set of developments at early April triggers synchronized and policy derived reactions across connected markets. At the same time, the regional contrasts embedded in the three subfigures of Figure \ref{fig04} are equally instructive. For example, the European panel (see Figure \ref{fig04} (B)) typically shows a plateau before the final fast increase (consistent with repricing along channels exposed to trade) whereas the Asian panel (see Figure \ref{fig04} (C)) often exhibits a quicker partial normalization (consistent with a faster digestion of the initial shock or a greater mix of offsetting local flows). Importantly, even in markets where entropy increases over the post-inauguration period, such as Bovespa in Table \ref{tab03}, the cumulative entropy still shows short intervals with sharp increases. This indicates that the market experiences distinct and well localized extreme episodes. This underscores that cumulative entropy complements level comparisons. It is sensitive to the timing and intensity of information changes, not just to whether average informational complexity is higher or lower across long windows.

Together, it can be argued that these April subfigures convey two linked pictures. First, cumulative entropy employs the same idea of a structured volatility as developed earlier. This is to say, markets can remain volatile while the set of resulting patterns becomes more ordered and the cumulative statistic shows when that ordering becomes most acute. Second, by presenting the Americas, Europe, as well as Asia and Oceania side-by-side (see Figures \ref{fig04} (A)-(C)), the figures show how a global narrative can yield a broadly synchronous increase while still allowing for a local heterogeneity in slope, peak, and unwind dynamics. In other words, cumulative entropy reveals when market dislocations occur and how concentrated they are, whereas traditional dispersion measures indicate only how large those dislocations are. The April~1–11 sequence in Figures \ref{fig04} (A)-(C) thus provides a compact and high resolution view of extreme events.

\begin{figure}[ht!]
\includegraphics[width=0.5\columnwidth]{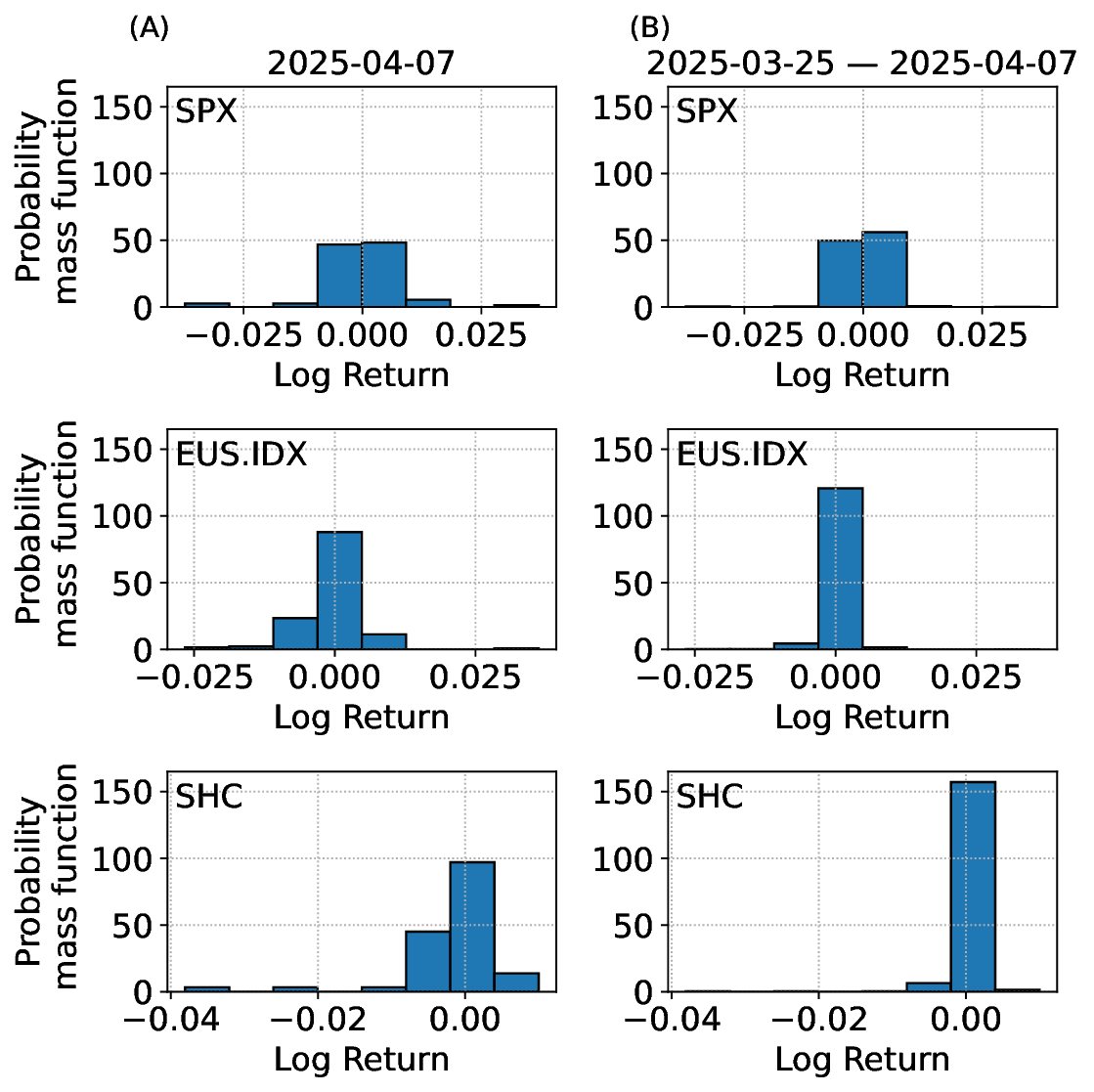}
\caption{The discrete probability mass function for the 5 minutes data, for April 7th, 2025 (A) and March 25th - April 7th, 2025 (B) across representative Americas, European as well as Asian and Oceania stock indices.}
\label{fig05}
\end{figure}

In order to give a further intuitive understanding of how cumulative entropy highlights extreme events or shocks, we plot the return distribution snapshots around representative points in time in Figure \ref{fig05}. We notice that pronounced exponential signatures of extreme events are particularity well visible on April 7 for Americas, Europe as well as Asia and Oceania. Of particular interest is the behavior of SPX, EUROSTOXX50 and SHC indices, respectively. Again, the observed behavior is attributed to the emergence of an extreme volatility in global stock markets due to the introduced tariffs. However, these results are not only representative to the core functionality of the cumulative entropy but also allow a fundamental insight into its underlying mechanism. In particular, Figure \ref{fig05} presents the distributions of the aforementioned indices corresponding to the extreme event and to the broader sample that includes that event together with the preceding 14 days. As we can see, the single day distributions (see Figure \ref{fig05} (A)) are relativity flat (which implies a higher entropy), particularly in terms of the EUROSTOXX50 and SHC indices, which exhibit a more pronounced exponential behavior than SPX (see Figure \ref{fig04} for more details). On the other hand, the collective distributions (see Figure \ref{fig05} (B)) are more peaked and, hence, concentrate around particular outcomes (which implies a lower entropy). This follows the signature behavior of cumulative entropy and a related interpretation that an extreme day should be viewed as the most chaotic/disturbed one where the preceding data has a more ordered character.

In sum, the obtained results show that the first 100 days were impactful for prices in a manner that is measurable, globally resonant, and conceptually unified. This corresponds to an elevated dispersion without a commensurate informational diffusion, accompanied by well-localized and repeatable extreme events. Together, the entropy and dispersion results demonstrate that volatility alone is insufficient to characterize market regimes during this period. Instead, the information structure plays an essential role. The method advanced here, which combines entropy, dispersion, and cumulative entropy, provides a transparent framework for monitoring not only how strongly markets move, but also how constrained and organized those movements are in informational terms. This dual perspective enables a finer assessment of whether early patterns persist, weaken, or reorganize as policy signals evolve and macroeconomic feedbacks unfold. While longer horizons are required to assess durability, the present analysis establishes the near-term informational structure of market responses with precision.

Crucially, these findings open practical pathways for stochastic forecasting under turbulent conditions. The observed entropy compression implies that return dynamics deviate from maximally random diffusion and instead occupy reduced, structured state spaces. In such regimes, entropy and cumulative entropy naturally increase, acting as informational constraints that can shape short-horizon scenario sets by limiting trajectories to those consistent with a heightened deterministic organization. In this context, the already mentioned entropy-corrected geometric Brownian motion framework \cite{gupta2024} is particularly well suited. It preserves the analytical tractability of standard geometric Brownian motion while explicitly conditioning forecast ensembles on the empirically observed entropy state of the system. When combined with cumulative entropy timing signals, the corrected model allows forecast weights to adapt dynamically as regimes emerge, strengthen, or unwind, thereby forming a concrete bridge between a high-frequency diagnostic insight and deployable forecasting methodologies.

\section{Summary and Conclusion}
\label{sec:conclusion}

In summary, the evidence from dispersion, information complexity, and cumulative entropy yields a coherent characterization of a concentrated policy communication period, represented by the first 100 days following the presidential inauguration of Donald Trump on January~20th,~2025. First, the obtained results here show that markets considered during that time were volatile yet narratively constrained. The side-by-side analysis conducted here exposed that while standard deviation remained relatively high or unchanged across many indices, the entropy frequently declined. This was interpreted as an indication that large price changes were channeled through a narrow set of outcomes. Such decoupling also resolved the intuitive but incorrect expectation that high volatility must entail high entropy. Instead, a regime was highlighted in which repeating responses concentrate the probability mass into fewer return bins. To this end, the described discussion was supplemented by the cumulative entropy analysis, which allowed us to trace extreme events via entropic signatures and pinpoint when the aforementioned ordering tightens the most. We note that these signatures were not cosmetic; instead, they reflected notable changes in the information theory measures, providing a consistent platform to study the impact of political decisions within the considered time window.

On the question of methodology itself, the obtained results showed that information theory measures may be effective in characterizing financial data under concentrated policy communication. In particular, they allowed for observations that are not accessible through the analysis based on dispersion (standard deviation) alone. In particular, the cross-regional analysis supported a unified interpretation of a global market response with local nuances. It was also shown that the indices most exposed to trade and financial links exhibit a substantial entropy compression and simultaneous increases in cumulative entropy. In our opinion, that suggests a rapid propagation of a common narrative. At the same time, the character of the mentioned increases varied across indices, in agreement with local factors that modulate the global signal. Even markets that displayed a higher average entropy in the post-inauguration period still registered well-localized cumulative entropy maxima, confirming that discrete extreme events or shocks were widely felt.

Interpreted through this lens, it is argued here that the discussed early policy strategy may be viewed as a forward guidance shock. In other words, we understand it as a concentrated sequence of decisions or communications strong enough to produce large, thematically coherent global moves across various indices. In this structure, the United States appears to play the role of a pivotal information node, suggesting a strong projection of influence. Still, this should not be read as a strong judgment about the geopolitical power of the United States. Instead, the obtained results show that initial actions matter strongly because markets price expectations continuously.

\section*{Acknowledgments}

All Authors except R.G., M.W.J., A.G. and J.G., would like to acknowledge financial support under European Funds (project no. FESL.01.02-IP.01-0942/24).

\newpage
\appendix
\section{Summary Statistics}

This appendix provides complementary descriptive statistics for the financial indices analyzed in the main text. Table~\ref{tab02} reports summary statistics for both daily and intraday (5-minute) returns computed over a 100-day window surrounding the inauguration of Donald Trump on January 20th, 2025. The reported measures include the return count, mean, minimum, maximum, skewness, and kurtosis, offering a comprehensive overview of the distributional properties of returns across markets.

Statistics reported outside parentheses correspond to daily returns, while values in parentheses refer to intraday returns. To ensure consistency and comparability, all statistics are rounded to three decimal places, with the exception of the mean, which is expressed in scientific notation. Percentage differences used in the subsequent analyses are computed as, $(x_2 - x_1)/[(x_1 + x_2)/2]$.

\begin{table*}[h]
\renewcommand{\arraystretch}{1.9}
\setlength{\tabcolsep}{2.8pt}
\caption{The summary statistics of intraday and daily returns considered in this study. Both daily and intraday 5 minutes (in brackets) results are presented.}
\begin{tabular}{|p{2cm}|p{2cm}|p{2.5cm}|p{2.5cm}|p{2.5cm}|p{2.5cm}|p{2.5cm}|p{2.5cm}}
\hline
\textbf{Name index} & \textbf{Returns} & \textbf{Mean} & \textbf{Minimum} & \textbf{Maximum} & \textbf{Skewness} & \textbf{Kurtosis} \\ \hline \hline

S$\&$P 500  
& 69 \newline (5456) 
& $-1.199 \times 10^{-3}$ \newline $(-1.356 \times 10^{-5})$ 
& -0.062 \newline (-0.037) 
& 0.091 \newline (0.037) 
& 0.904 \newline (-0.680) 
& 8.287 \newline (92.017) \\

\hline
DJIA  
& 69 \newline (5460) 
& $-1.149 \times 10^{-3}$ \newline $(-1.227 \times 10^{-5})$ 
& -0.057 \newline (-0.034) 
& 0.076 \newline (0.034) 
& 0.766 \newline (-0.167) 
& 8.039 \newline (101.386) \\

\hline
NASDAQ Composite  
& 69 \newline (5460) 
& $-1.802 \times 10^{-3}$ \newline $(-2.160 \times 10^{-5})$ 
& -0.062 \newline (-0.042) 
& 0.115 \newline (0.039) 
& 1.137 \newline (-1.158) 
& 7.369 \newline (77.479) \\

\hline
Bovespa 
& 68 \newline (6126) 
& $1.394 \times 10^{-3}$ \newline $(1.614 \times 10^{-5})$ 
& -0.030 \newline (-0.019) 
& 0.031 \newline (0.020) 
& 0.305 \newline (0.536) 
& 0.655 \newline (57.067) \\

\hline
S\&P\slash TSX Composite 
& 70 \newline (5538) 
& $-0.188 \times 10^{-3}$ \newline $(-0.164 \times 10^{-5})$
& -0.048 \newline (-0.031) 
& 0.053 \newline (0.025) 
& -0.218 \newline (-3.915) 
& 4.179 \newline (118.913) \\

\hline
Euro Stoxx 50 
& 70 \newline (11592) 
& $0.042 \times 10^{-3}$ \newline $(0.056 \times 10^{-5})$ 
& -0.056 \newline (-0.027) 
& 0.083 \newline (0.036) 
& 0.798 \newline (1.620) 
& 9.001 \newline (129.027) \\

\hline
FTSE 100 
& 85 \newline (18897) 
& $-0.006 \times 10^{-3}$ \newline $(0.011 \times 10^{-5})$ 
& -0.057 \newline (-0.030) 
& 0.060 \newline (0.021) 
& -0.042 \newline (-2.205) 
& 9.823 \newline (192.786) \\

\hline
DAX 
& 70 \newline (7228) 
& $0.990 \times 10^{-3}$ \newline $(0.993 \times 10^{-5})$ 
& -0.051 \newline (-0.111) 
& 0.044 \newline (0.078) 
& -0.453 \newline (-10.801) 
& 1.252 \newline (1075.266) \\

\hline
WIG20 
& 70 \newline (6659) 
& $2.155 \times 10^{-3}$ \newline $(2.320 \times 10^{-5})$ 
& -0.066 \newline (-0.078) 
& 0.044 \newline (0.061) 
& -0.643 \newline (-4.502) 
& 1.354 \newline (466.542) \\

\hline
Nikkei 225 
& 68 \newline (4151) 
& $-1.122 \times 10^{-3}$ \newline $(-1.557 \times 10^{-5})$ 
& -0.082 \newline (-0.037) 
& 0.087 \newline (0.037) 
& 0.289 \newline (-1.948) 
& 6.560 \newline (99.541) \\

\hline
Shanghai Composite 
& 65 \newline (3169) 
& $0.163 \times 10^{-3}$ \newline $(0.360 \times 10^{-5})$ 
& -0.076 \newline (-0.038) 
& 0.018 \newline (0.010) 
& -4.335 \newline (-7.594) 
& 26.526 \newline (197.775) \\

\hline
Hang Seng Index  
& 66 \newline (4387) 
& $1.582 \times 10^{-3}$ \newline $(2.775 \times 10^{-5})$ 
& 0.002 \newline (-0.087) 
& -0.142 \newline (0.029) 
& 0.039 \newline (-8.193) 
& 17.892 \newline (294.748) \\

\hline
Nifty 50 
& 67 \newline (5014) 
& $0.620 \times 10^{-3}$ \newline $(0.910 \times 10^{-5})$ 
& -0.033 \newline (-0.046) 
& 0.022 \newline (0.019) 
& -0.400 \newline (-9.994) 
& 1.219 \newline (455.761) \\

\hline
S\&P\slash ASX 200 
& 83 \newline (16804) 
& $-0.455 \times 10^{-3}$ \newline $(-0.160 \times 10^{-5})$ 
& -0.051 \newline (-0.024) 
& 0.066 \newline (0.017) 
& 0.643 \newline (-2.167) 
& 8.957 \newline (102.081) \\

\hline
NZX 50 
& 68 \newline (4310) 
& $-1.396 \times 10^{-3}$ \newline $(-2.061 \times 10^{-5})$ 
& -0.037 \newline (-0.020) 
& 0.033 \newline (0.033) 
& -0.377 \newline (3.645) 
& 3.557 \newline (261.817) \\

\hline \hline
\end{tabular}

\label{tab02}
\end{table*}

\begin{table*}[h]
\renewcommand{\arraystretch}{1.9}
\setlength{\tabcolsep}{2.8pt}
\caption{Entropy measures before and after the inauguration of Donald Trump on January 20th, 2025. The statistics are reported for the 100-day window bracketing presidential inauguration.}
\begin{tabular}{|p{3.3cm}|p{2.2cm}|p{2.2cm}|p{2.2cm}|p{2.2cm}|p{2.2cm}|p{2.2cm}|}
\hline
\textbf{Name index} & \textbf{Entropy \newline before} & \textbf{Entropy \newline after} & \textbf{Percentage Difference} & \textbf{St. dev. \newline before} & \textbf{St. dev. \newline after} & \textbf{Percentage Difference} \\
\hline\hline

S$\&$P 500  & 1.465 & 1.099 & 28.57\% & 0.008 & 0.008 & 3.68\% \\
\hline
Dow Jones Industrial Average  & 1.177 & 1.161 & 1.36\% & 0.008 & 0.009 & 3.51\% \\
\hline
NASDAQ Composite &   1.565 & 1.107 & 34.32\% & 0.009 & 0.008 & 12.20\% \\
\hline
Bovespa   & 1.534 & 1.660 & 7.92\% & 0.009 & 0.011 & 20.29\% \\
\hline
S\&P\slash TSX Composite   & 1.482 & 1.200 & 21.02\% & 0.011 & 0.007 & 39.57\% \\
\hline
Euro Stoxx 50 &  1.655 & 1.058 & 44.05\% & 0.009 & 0.009 & 4.49\% \\
\hline
FTSE 100  & 1.758 & 0.893 & 65.20\% & 0.009 & 0.009 & 3.43\% \\
\hline
DAX Index  & 1.693 & 1.573 & 7.33\% & 0.009 & 0.009 & 1.12\% \\
\hline
WIG20  & 1.654 & 1.574 & 4.96\% & 0.008 & 0.008 & 2.50\% \\
\hline
Nikkei 225  & 1.666 & 1.044 & 45.94\% & 0.009 & 0.009 & 7.82\% \\
\hline
Shanghai Composite Index   & 1.583 & 0.852 & 60.02\% & 0.009 & 0.007 & 17.28\% \\
\hline
Hang Seng Index   & 1.640 & 1.048 & 44.03\% & 0.009 & 0.007 & 20.00\% \\
\hline
Nifty 50  & 1.672 & 1.556 & 7.18\% & 0.009 & 0.012 & 25.49\% \\
\hline
S\&P\slash ASX 200   & 1.400 & 1.037 & 29.81\% & 0.009 & 0.008 & 8.38\% \\
\hline
NZX 50  & 1.690 & 1.303 & 25.89\% & 0.012 & 0.012 & 0.85\% \\
\hline\hline
\end{tabular}
\label{tab03}
\end{table*}

\newpage
\section{Cumulative entropy}

Appendix~B presents the cumulative entropy analysis for the financial indices considered for the 100-day time window after the inauguration of Donald Trump on January~20th,~2025. The figures illustrate the temporal evolution of cumulative entropy computed from the return distributions, allowing for a dynamic assessment of information content and market response. Results are shown separately for American, European as well as Asian and Oceania markets in order to highlight potential regional differences. Note that each set of the results is divided into two subsets of figures.

The cumulative entropy is obtained by aggregating entropy measures over time, providing a smoothed representation of changes in market dynamics. Each figure reports the temporal evolution of cumulative entropy for selected representative indices within a given region. This appendix complements the results discussed in the main text by offering an additional insight into the persistence and magnitude of entropy variations across markets.

\begin{figure}[h]
\centering
\includegraphics[scale=0.75]{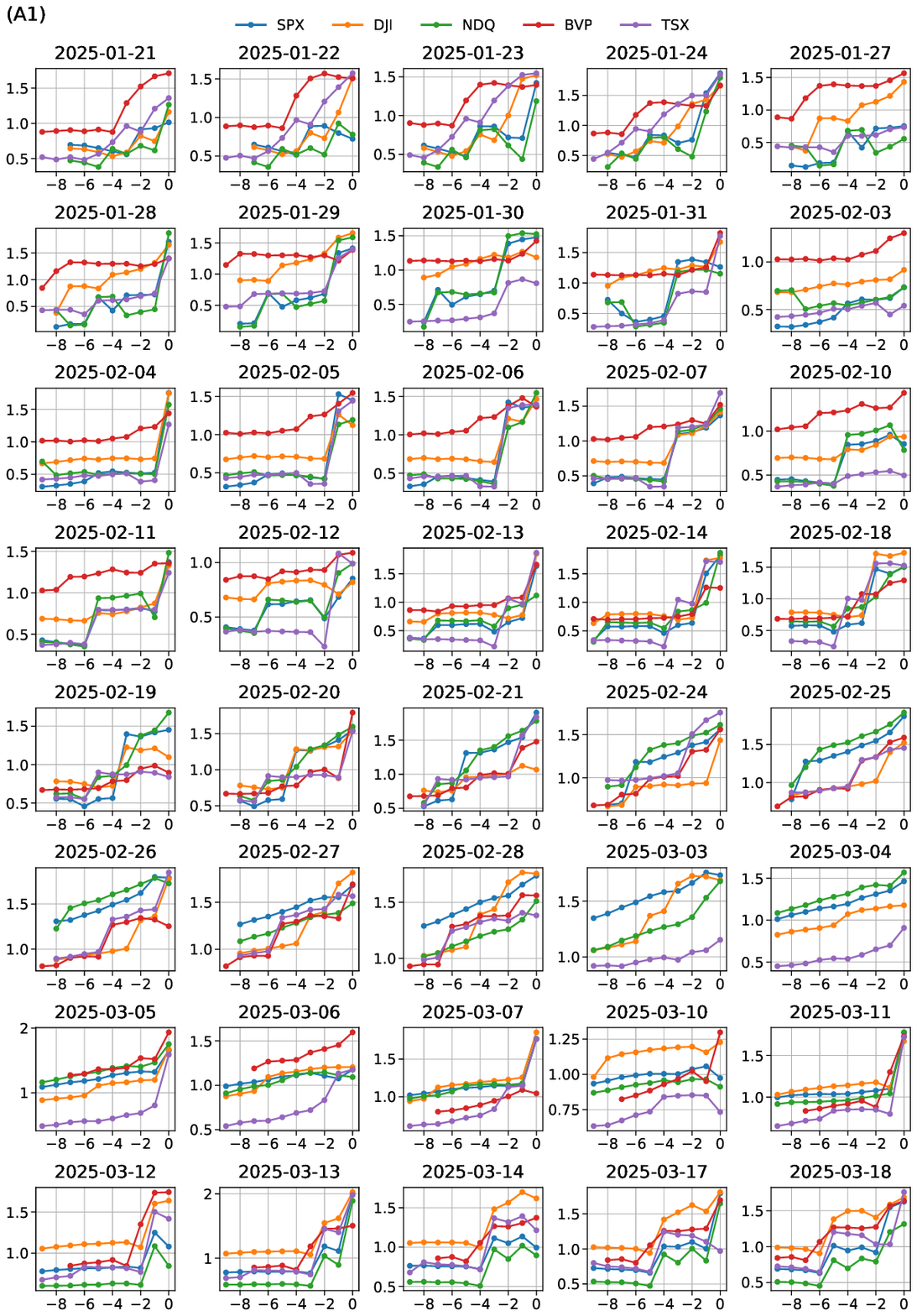}
\caption{Cumulative entropy for January~21--March~18, 2025 across representative Americas stock indices (subset~1).}
\label{fig05(1)}
\end{figure}

\begin{figure}[h]
\centering
\includegraphics[scale=0.75]{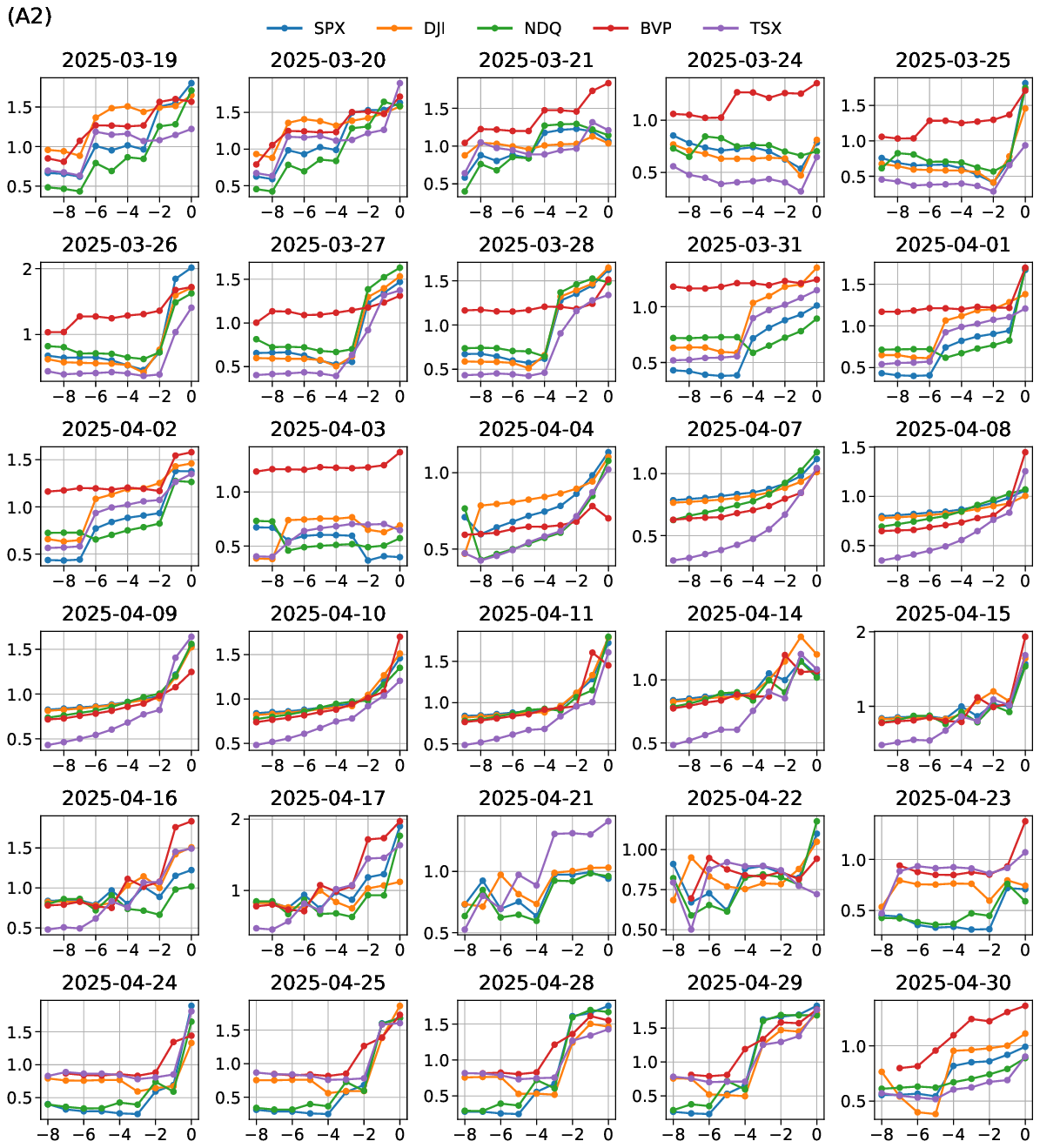}
\caption{Cumulative entropy for March~19--April~30, 2025 across representative Americas stock indices (subset~2).}
\label{fig05(2)}
\end{figure}

\begin{figure}[h]
\centering
\includegraphics[scale=0.75]{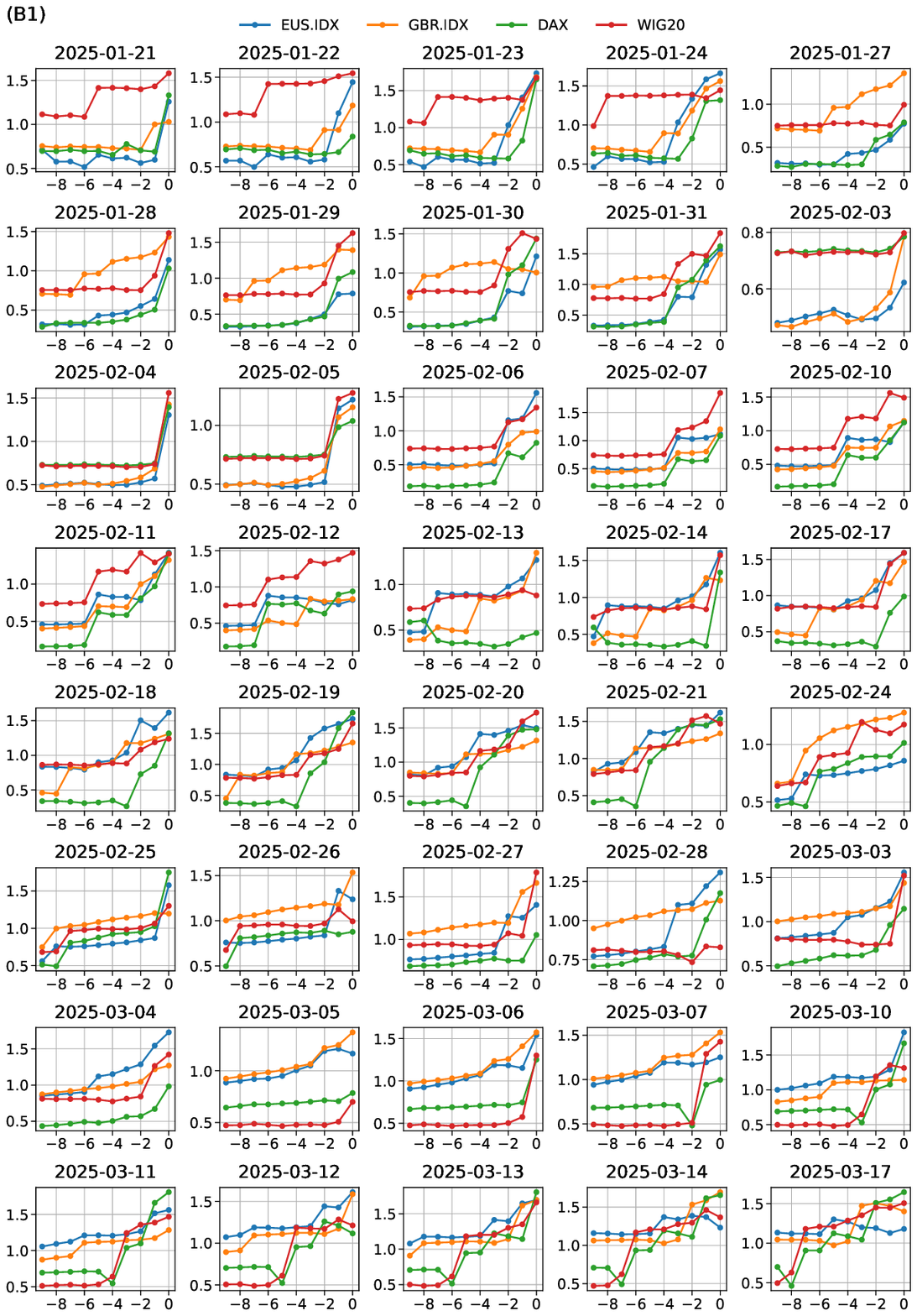}
\caption{Cumulative entropy for January~21--March~17, 2025 across representative European stock indices (subset~1).}
\label{fig06 (1)}
\end{figure}

\begin{figure}[h]
\centering
\includegraphics[scale=0.75]{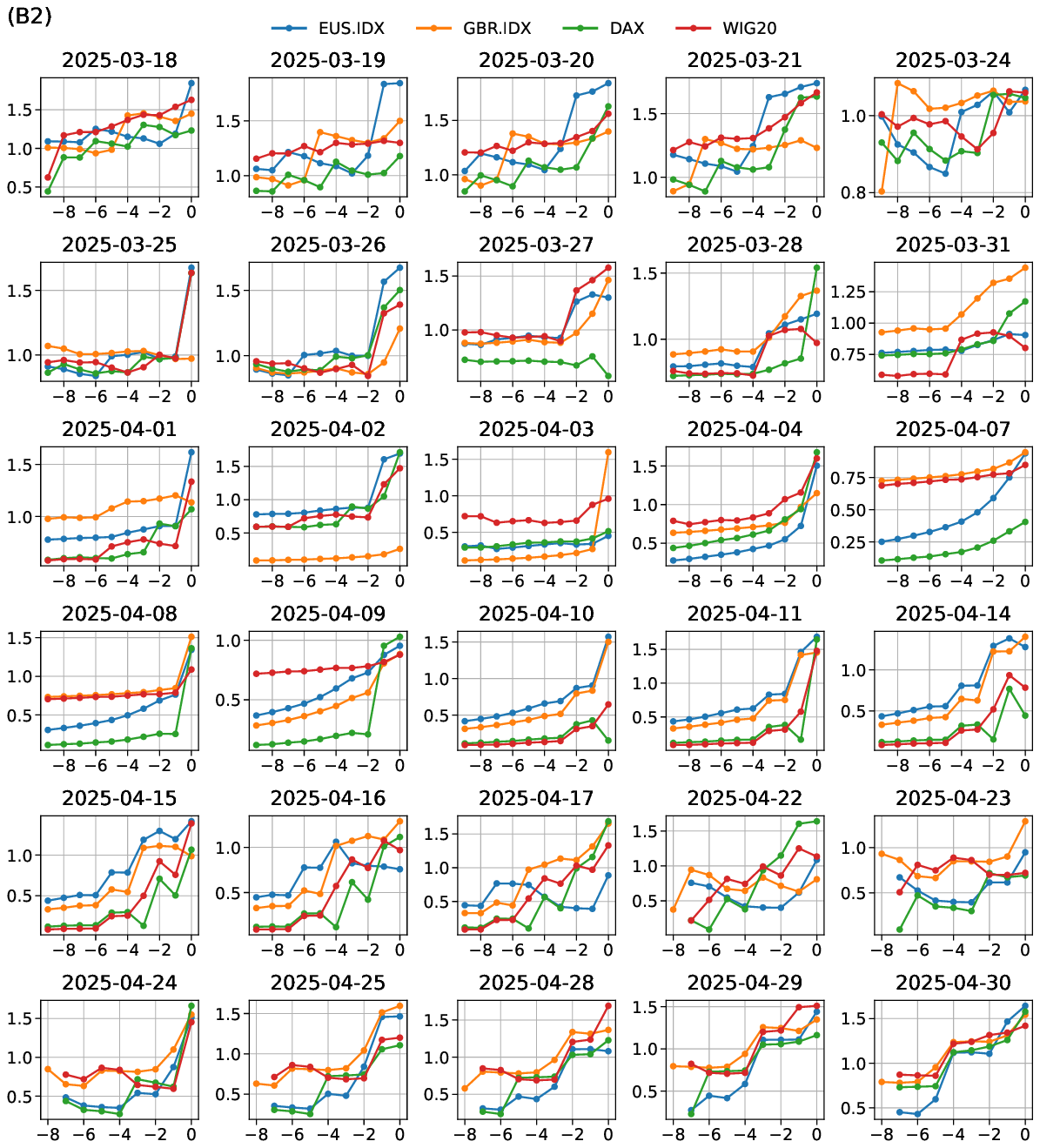}
\caption{Cumulative entropy for March~18--April~30, 2025 across representative European stock indices (subset~2).}
\label{fig06 (2)}
\end{figure}

\begin{figure}[h]
\centering
\includegraphics[scale=0.75]{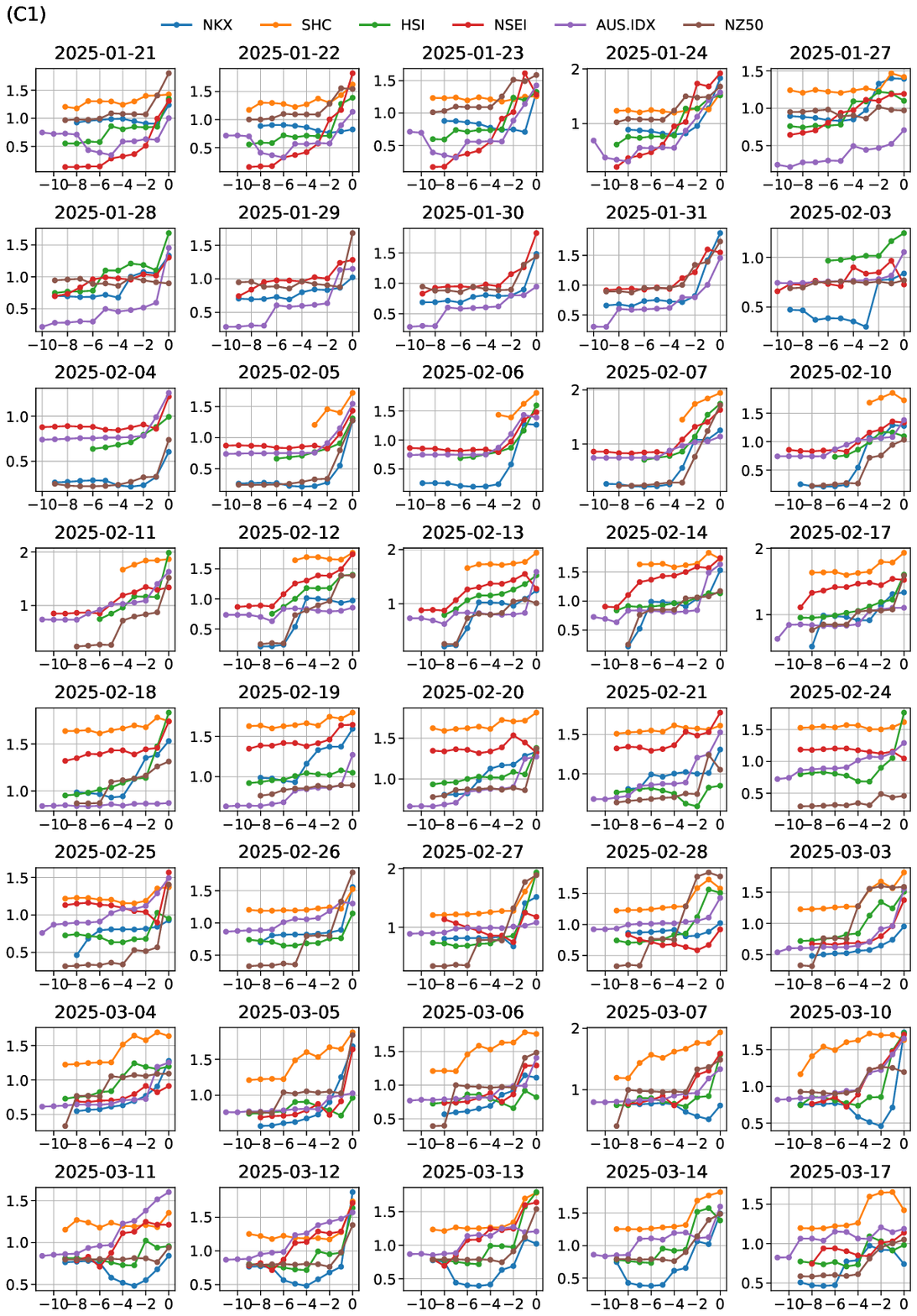}
\caption{Cumulative entropy for January~21--March~17, 2025 across representative Asian and Oceania stock indices (subset~1).}
\label{fig07 (1)}
\end{figure}

\begin{figure}[h]
\centering
\includegraphics[scale=0.75]{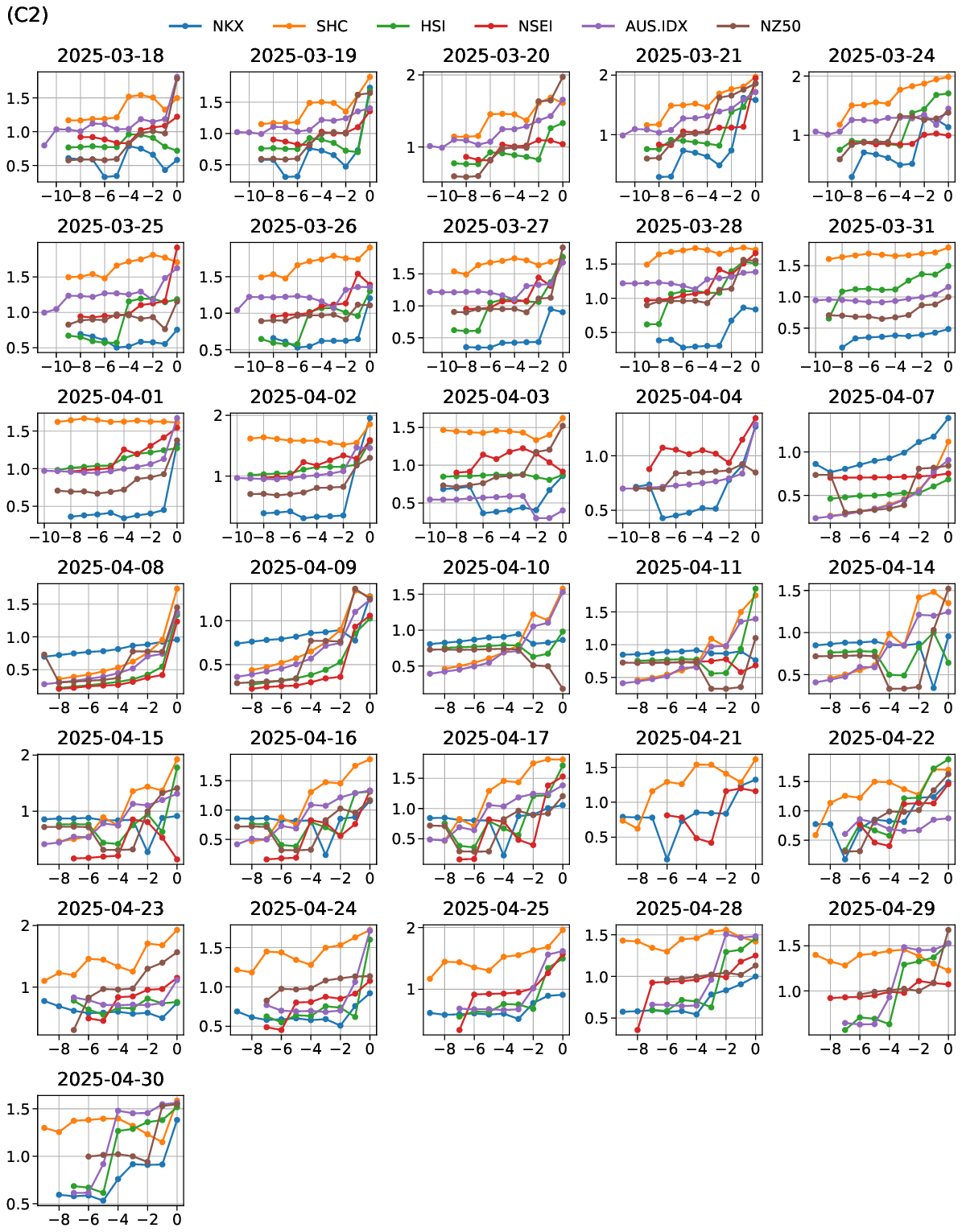}
\caption{Cumulative entropy for March~18--April~30, 2025 across representative Asian and Oceania stock indices (subset~2).}
\label{fig07 (2)}
\end{figure}

\bibliography{manuscript}
\end{document}